# Maximizing switching current of superconductor nanowires via improved impedance matching

Labao Zhang, Xiachao Yan, Xiaoqing Jia, Jian Chen, Lin Kang, and Peiheng Wu

School of Electronic Science and Engineering, Nanjing University, Nanjing, 210093, China

## ABSTRACT

The temporary resistance triggered by phase slip will result in the switching of superconductor nanowire to a permanent normal state, decreasing the switching current. In this letter, we propose an improved impedance matching circuit that releases the transition triggered by phase slips to the load resistor through the RF port of a bias tee. The transportation properties with different load resistors indicate that the switching current decreases due to the reflection caused by impedance mismatching, and is maximized by optimized impedance matching. Compared to the same setup without impedance matching circuit, the switching current was increased from 8.0 μA to 12.2 μA in a niobium nitride nanowire after releasing the temporary transition triggered by phase slips. The leakage process with impedance matching outputs a voltage pulse which enables the user to directly register the transition triggered by phase slips. The technique for maximizing the switching current has potential practical application in superconductor devices, and the technique for counting phase slips may be applied to explore the quantum behavior of phase slips.

[Insert Physics and Astronomy Classificaiton Scheme Codes]

Though it is commonly assumed that the electrical resistance and expulsion of magnetic flux fields are zero in superconductors, the resistance of a superconductor nanowire is never exactly zero due to the fact that phase slips have a nonzero probability at any finite temperature. Superconductor nanowire phase slip phenomena[1] have been intensively investigated over the last decade in effort to better understand the superconductor mechanism [2-6] and to extend its application in ultra-sensitive electronic devices [7-12].

Phase slips can not only be observed in one-dimensional superconductors, but also in quasi-2-dimensional structures [13, 14]. Generally, phase slips include the thermal-activated phase slip (TAPS)[15], thermal barrier-crossing by the order-parameter field at high temperatures, and quantum phase slip (QPS)[3], a quantum tunneling event occurring at low temperatures. Recently, more complex phenomena such as quantum phase slip pairs[16] and coherent quantum phase slips[17]have been reported in superconductor nanowires. Phase slip transition is unfortunately difficult to measure as it is random, weak, and fast. For this reason, the recognition of phase slips has heavily relied on indirect evidence such as altered transportation properties[6] and the resistance-temperature relationship[3].

Previous researchers have utilized long nanowires with uniform cross-sections and used them to explore the properties of superconductor nanowires in terms of collective effects [3, 18]. Generally speaking, ideal superconductor nanowires should be in a state of zero resistance when cooled below their transition temperature and below their critical current. In practice, however, the superconductor nanowire exhibits a resistance tail [3]and switching current suppression[19] related to phase slip phenomena. The increased length of superconductor nanowire amplifies the superconducting transition over the nanowire, resulting in a measurable resistance even far below the transition temperature.

Another indirect evidence [20] was obtained from current-voltage (*I-V*) curves, where it is possible to deduce a switching current for superconductor nanowire transition to the resistive state by phase slips. Many previous researchers have reported that an increasing number of phase slips with increased segment length produces observable differences in the *I-V* curves. The switching current is apparently smaller than the practical critical current[19]. A suppressed critical current can also be considered to indicate the transition in the superconductor nanowire current to the resistive state, as evidenced by the *I-V* curve, during the sweeping-up current. When the nanowire is biased with a current higher than the switching current, the superconductor nanowire is triggered to transition from the superconducting state to a permanent normal state. The measured transition current is called the "switching current" in the *I-V* curve. When the switching current is lower than the critical current, this transition is mainly activated by self-heating in the resistive area as-triggered by phase slips. Brenner et al. [21] reported that the switching current of superconducting nanowires can be increased with an external resistor shunt. For practical devices, such as superconductor nanowire single photon detectors, the superconductor nanowire switching current must be as high as possible.

In this study, we built a circuit with a bias tee and optimized load impedance designed to maximize the switching current. The proposed circuit can counter the transition triggered by the phase slip to prevent the superconductor nanowire from switching to the normal state when the bias current is lower than its critical current. We release the transition to enable nanowire recovery from a temporary resistive state by releasing the bias current. The switching current was increased from 8.0μAto 12.2μA after the proposed modification.

A meandered niobium nitride (NbN) nanowire as shown in Fig. 1(a) with parallel wires connected in series was fabricated via sputtering, electron beam lithography (EBL), and reactive ion etching (RIE).The designed structure has a total length of 2 mm and covered an area of 20 μm × 20 μm. We selected this structure for several reasons: First, the increased length of the nanowire enhanced the appearance of phase slip over the nanowire; second, this structure yields a uniform cross-section for high-quality film in a limited area and with favorable micro-fabrication process parameters; third, this uniform structure minimizes the impact of imperfections such as weak links, film defects, or chip yields.

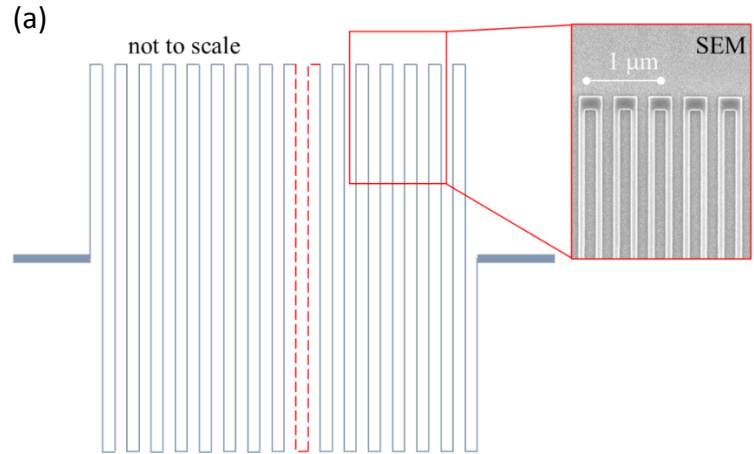

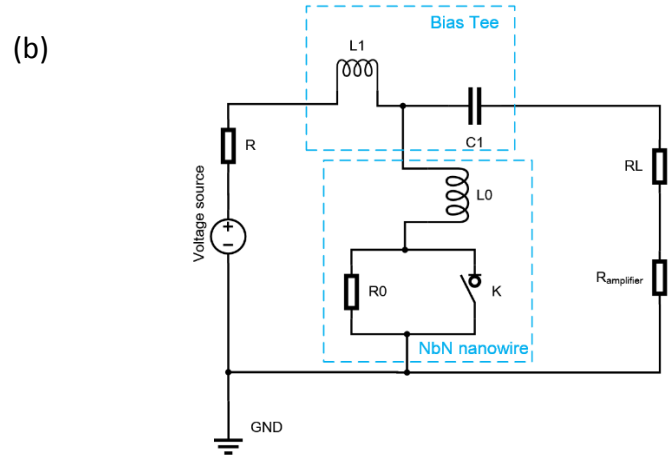

**Fig. 1** (a) Designed nanowire structure with a total length of 2 mm, area of 20 μm ×20 μm (insert: SEM image of patterned NbN nanowire); (b) Circuit for transportation measurement. The equivalent circuits of NbN nanowire and Bias Tee were shown in the rectangular frames with blue broken boundary. $R_{amplifier}$ is the impedance of amplifier, and RL is a fixed resistor series connected in the circuit.

The NbN films were deposited on a single-crystal magnesium fluoride substrate under optimized conditions. The NbN films have a thickness of 5 nm, a superconductor transition temperature ($T_c$) of 7.1 K, and a sheet resistance of $R_s$ = 320 Ω/square. The designed structure was patterned by 100 keV EBL (EBPG 5200) with a dose of 600 μC/cm$^2$ on a positive resist AR-P 6200.04 (Allresist). The patterned structure was transferred to the NbN films by RIE with a mixer of SF$_6$ and CHF$_3$. The width of the nanowire is 80 nm with a 120 nm inter-distance, as shown in the insert of the SEM image in Fig. 1(a). For impedance matching in the RF band, the nanowire was connected to a 50-ohm coplanar waveguide fabricated on the substrate. The coplanar waveguide was connected to the bias circuit at room temperature through a coaxial cable (DC-6 GHz).

The circuit settings are shown in Fig. 1(b). A programmable voltage source was adopted to bias the superconductor nanowire after being connected in series with a fixed resistor($R$=100kΩ). A bias tee was introduced in the circuit, as the DC source was connected to the NbN nanowire through the DC port with an inductor L1. The inductor blocked the RF noise from the environment, and the RF signal from the nanowire was also filtered. The RF port of the bias tee, with a band of 0.1-4200 MHz, was connected to a low noise amplifier with an impedance of 50 Ω and a gain of 51 dB in the 1kHz-1GHz, which covers most energy of the output pulse triggered by phase slips in this experiment. An additional resistor RL was series-connected with the amplifier to analyze the impact of impedance matching on switching current. All readout circuits, including bias tee, RL and amplifier, were installed at room temperature. A closed-cycle system based on a G-M cryo-cooler was used in this experiment. The NbN nanowire was mounted on the cold head of the cryo-cooler with set temperature of 2.5 K and fluctuation below 5 mK.

(a)

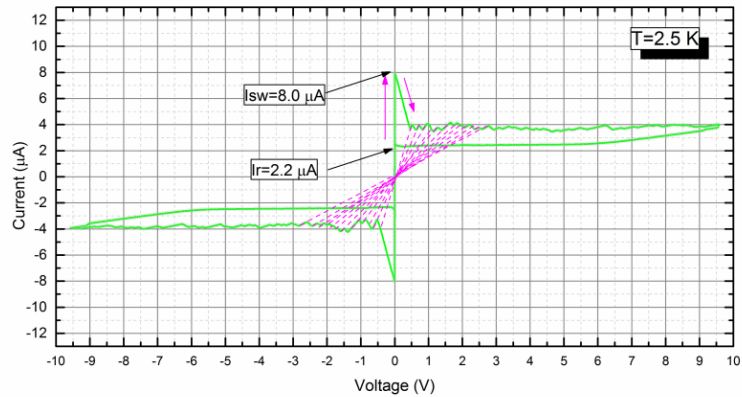

(b)

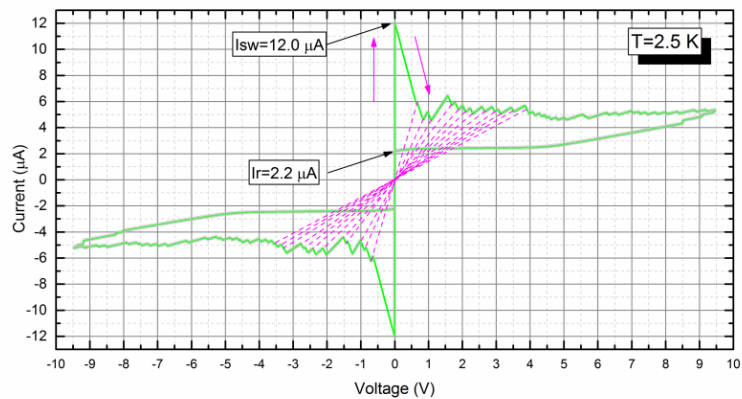

**Fig. 2** Typical *I-V* curves at 2.5 K measured (a) without impedance matching circuit and (b) with the improved impedance matching circuit. Note: Pink arrows indicate the sweeping sequence; pink dashes mark the resistive branches.

The typical *I-V* curves of the NbN nanowire measured at 2.5 K are shown in Fig. 2, where the voltage source (connected with a 100k$\Omega$ resistor) can be changed from low to high as necessary. For comparison, we show two *I-V* curves: One without the bias tee and RL in Fig. 2(a) and one with bias tee and optimized RL in Fig. 2(b).The nanowire exhibited a superconducting state along the zero-voltage line in both curves and abruptly transited to the resistive state over the negative resistive line, where the nanowire was partially in the normal state. In Fig. 2, neither $I$-$V$ curve contains a residual voltage tail at any current lower than the switching current; this reflects the high quality of the fabricated nanowire. The branches in the $I$-$V$ curves correspond to the resistances of the meandered nanowires. As the voltage was swept from high to low, the nanowire was able to recover to the superconducting state only when the current was less than $I_r$, which is referred to as the "retrapping current" [18]. The retrapping current is the same in both curves, but the switching current is lower in the curve shown in Fig. 2(a) than in Fig.2 (b). We attribute this difference to leakage of the transition triggered by the phase slips. To analyze the mechanism of transition leakage, we divided the transitions from superconducting state to normal state into two types: Temporary transition and permanent transition.

Generally speaking, a superconductor should transition to the normal state permanently if $T>Tc$, $I>Ic$, or $H>Hc$. Our superconductor nanowire transitioned to the resistive state, however, when the bias current was lower than its critical current (Fig. 2(a)). The probability of phase slips, including QPS and TAPS, increased exponentially during the $I$-$V$ measurement period when the bias current was increased. A temporary transition was triggered by phase slips. The nanowire was then heated by Joule heating $I^2R$, which increased the resistive area which produced even more heating. This cycle continued until the superconducting state over the nanowire was destroyed. The self-heating effect forced the superconductor to return to the normal state permanently even when the bias current was less than the critical current, then the temporary transition led to a permanent transition.

As shown in Fig. 2(b), an additional load RL was introduced to release the transition caused by phase slips. A previous simulation of a superconductor nanowire reported that the temperature of nanowire increases quickly and recovers slowly due to a sequence of phase slips[6]; the simulation indicated that the resetting time is about 0.2 ns. The temperature was increased 100mK in the peaks for each phase slip, even at the K level for multi-phase slips. As a result, the superconducting nanowire partially transited to normal state where the bias current exceeded its critical current. When a temporary resistance formed over the nanowire, most bias current streamed to the load resistor because the load impedance was much lower than the temporary resistance. Accordingly, the practical current in the nanowire was reduced and the temporary resistive area recovered to the superconducting state. The temporary transition was released via the additional load through the RF port resulting in a voltage pulse over the load. Thus, the nanowire can be biased higher using the improved circuit in Fig. 1(b) compared to the setup without the impedance matching circuit.

The leakage process was recorded by an oscilloscope after amplification as shown in Fig. 3. A typical voltage pulse is shown in Fig. 3(a) at $I_{sw}$=11 μA and without RL. The recovery time indicated in Fig. 3(a) is about 50 ns, which was much longer than the previously reported simulation result [6]. The pulse was mainly slowed by the kinetic inductance of the superconductor nanowire, which is equivalent to an inductor. The peak of the output pulse was ~454μV and the calculated value was 440 μV based on the stream current ($I_{sw}$-$I_r$)×50 ohm. Considering the fluctuation caused by the signal noise of 64 μV (peak-to-peak), the pulse height was consistent with our prediction based on stream current theory. All output pulses exhibited uniform height regardless of the phase slips, marking a departure from the previously published simulation results [6]. The uniform height indicated that the observed pulse profile was mainly affected by the circuit, although it was triggered by the phase slips. To this effect, we can calculate the phase slip rate/possibility by measuring the voltage pulse as shown in Fig. 3(a).

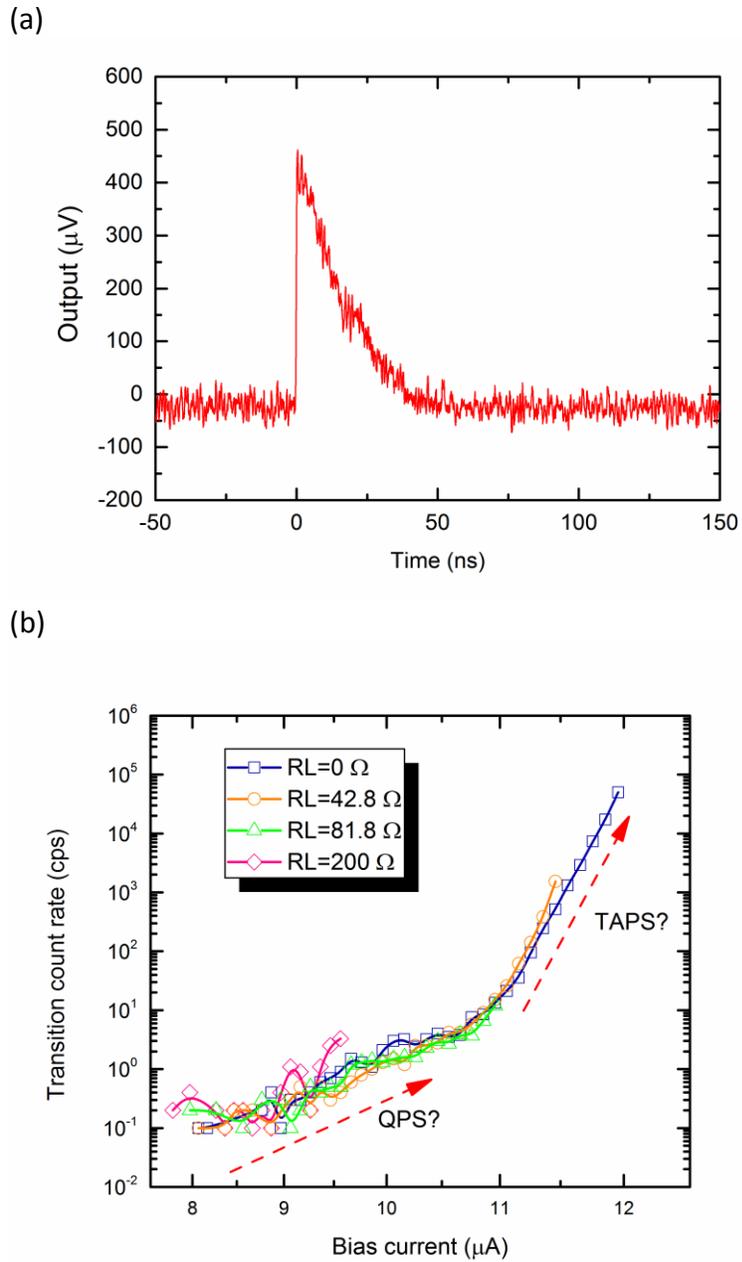

**Fig. 3** (a) Pulse profile triggered by phase slips over the load; (b) Count rate of output pulse versus bias current for different RL values

The pulse rates versus bias current are shown in Fig. 3(b).A digital counter was used to count the pulse rate by discriminating the rising edge; the counting rates were averaged over 10 seconds. Four values of the additional load RL (Fig. 1 (b)) were adopted, including 0, 42.8, 81.8, and 200 Ω. All the counting rates increased over the bias currents. The count rates were similar at any given bias current, though the switching currents changed in the circuit. This indicates that the proposed bias setting only changed the switching current. This observation supports our finding that the possibility of phase slippage, scaled as $\exp(-\Delta F(T, I)/kT)$, is determined by the temperature and the bias current of the nanowire. The switching current to normal state values differed across the different load RL values: 12 µA, 11.5 µA, 11 µA, and 9.7 µA correspond to RL=0, 42.8, 81.8, and 200 Ω, respectively. We speculate that the presence of the additional load changed the impedance matching at the RF port. This mismatch may have prevented the leakage of phase slips in the load port.

There are two areas with different slopes shown in Fig. 3(b). When the bias current was lower than 10.5 µA, the slope of the count rate was relatively low but increased abruptly when the bias current exceeded 10.5 µA. In theory, the count rate of QPS should be higher than the count rate of TAPS at low bias current, but here, it increased with a lower slope over the bias current – previous researchers made similar observations [6]. Figure 3(b) shows two regimes, one with the low count rate and low slope at low bias current and another with high count rate and high slope at high bias current. The regime with lower bias current may be QPS, and that with higher bias current may be TAPS according to features of phase slips and the profiles indicated in Fig. 3(b). However, the data at low bias current was rough due to low count rate. More evidence is yet needed to draw fully accurate conclusions.

The statistical values of switching currents for different RL resistances are presented in Fig. 4(a).In general, the switching current fluctuates with the random behavior of phase slips. Sahu et al.[6] argued that the distribution width is scaled with the thermal noise and therefore decreases as temperature decreases, and is saturated at low temperature where thermal fluctuations are frozen out and only quantum fluctuations remain. In our experiment, the distribution of switching currents was based on 200 repeated *I-V* measurements (open squares)at 2.5 K and fitted quite well to a Gaussian function (solid line).The width of the distribution was not strictly regular, as shown in Fig. 4(a), for small statistical samples measured with our programmable source, which exhibited low noise but was time-consuming.

Figure 4(b) presents the switching current over RL as-deduced from the data shown in Fig. 4(a). The switching current was 12 μA without RL, but increased slightly to 12.2 μA when RL increased to 13 Ω. After that point, the switching current decreased when RL was further increased. As described above, all components, including the amplifier, coplanar waveguide, and coaxial cable, have 50 Ω impedance. Ideally, the switching current is maximized with our improved matching when the RL=0 Ω. However, the peak was reached at RL=13 Ω (Fig. 4(b)).

Impedance matching can be quantified by reflection $[(R_{load}-r)/(R_{load}+r)]^2$, where $R_{load}$=RL+50 Ω and r is the internal impedance. We can determine the relationship of switching current versus reflection accordingly. Assuming r=64 Ω, the curve was a uniform function of reflection as shown in Fig. 4(c). The practical internal impedance includes the residential resistance of the superconductor nanowire and contact resistance, so it seems reasonable that the practical resistance was slightly higher than 50 Ω. This explanation of the results shown in Fig. 4(b) is also consistent with our first claim that improved matching maximizes the switching current via leakage of the phase slip transition.

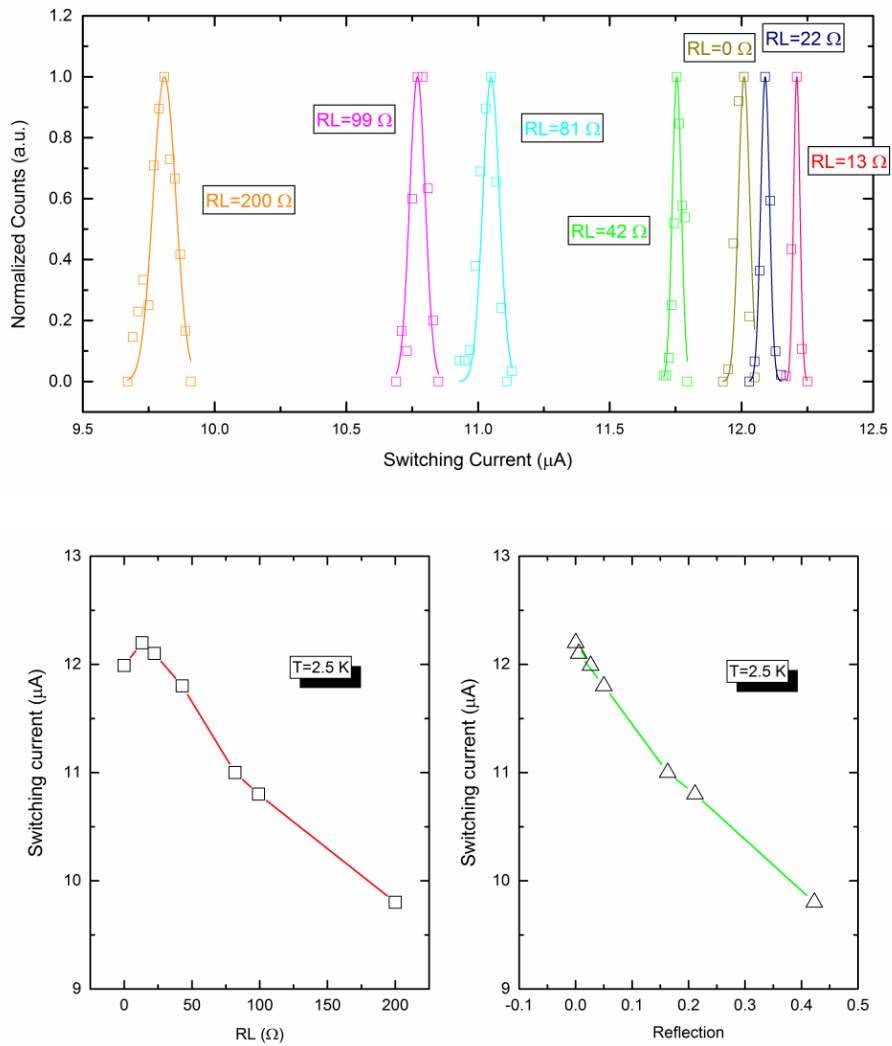

**Fig. 4** (a) Distribution of switching current deduced from *I-V* curves at 2.5K with different RL; (b) Switching current versus RL;(c) Switching current versus reflection assuming internal impedance at 64 Ω

In conclusion, the response pulse corresponding to the transition triggered by phase slippage was successfully released through an improved impedance matching circuit. The switching current of our nanowire increased from 8.0 μA to 12.2 μA at 2.5 K. The leakage process with impedance matching represents a novel approach to registering the transition triggered by phase slips directly, as well. Therefore, it may be applicable to further exploration of the quantum behavior of phase slips and to improve the performance superconducting devices.

[Insert Running title of <72 characters]

[Insert Running title of <72 characters]

[Insert Running title of <72 characters]